\documentclass[10pt]{iopart}
\usepackage{iopams}
\usepackage{graphicx,bm,cite}
\def\unitev{\,{\rm eV}}

\def\unitwn{\,{\rm cm}^{-1}}
\def\unitps{\,{\rm ps}}
\def\unitfs{\,{\rm fs}}

\begin{document}


\title{Selective coherent phonon mode generation in single wall carbon
  nanotubes}

\author{Ahmad.~R.~T.~Nugraha, Eddwi H.~Hasdeo, and Riichiro~Saito}
\address{Department of Physics, Tohoku University, Sendai 980-8578,
  Japan}
\ead{\mailto{nugraha@flex.phys.tohoku.ac.jp}}
\date{\today}
 

\begin{abstract}
  The pulse-train technique within ultrafast pump-probe spectroscopy
  is theoretically investigated to excite a specific coherent phonon
  mode while suppressing the other phonon modes generated in single
  wall carbon nanotubes (SWNTs).  In particular, we focus on the
  selectivity of the radial breathing mode (RBM) and the G-band for a
  given SWNT.  We find that if the repetition period of the pulse
  train matches with integer multiple of the RBM phonon period, the
  RBM amplitude could be kept while the amplitudes of the other modes
  are suppressed.  As for the G-band, when we apply a repetition rate
  of half-integer multiple of the RBM period, the RBM could be
  suppressed because of destructive interference, while the G-band
  still survives.  It is also possible to keep the G-band and suppress
  the RBM by applying a repetition rate that matches with integer
  multiple of the G-band phonon period.  However, in this case we have
  to use a large number of laser pulses having a property of ``magic
  ratio'' of the G-band and RBM periods.
\end{abstract}

\pacs{78.67.Ch,63.22.-m,73.22.-f,78.67.-n}


\maketitle
\ioptwocol

\section{Introduction}
Ultrashort laser pulses that are given to a single wall carbon
nanotube (SWNT) sample with duration less than a phonon period in the
SWNT may generate lattice oscillations coherently, which is usually
called as the coherent
phonons~\cite{dumi04,gambetta06-cp,lim06-cpexp,luer09-cp,makino09-cpdoping}.
Quantum mechanically, these oscillations can be treated using the
concept of coherent states, in which the amplitude-phase uncertainty
reach its minimum
value~\cite{zeiger92-cp,stanton94-cpmethod,hu96-qcp,merlin97-cp}.
Recently, by using sub-10-fs laser pulses, more than $10$ coherent
phonon modes with frequency around $100$--$3000\unitwn$ could be
observed in a (6,5) enriched single-chirality SWNT
sample~\cite{lim14-cpfund}.  This finding opened up possibilities to
utilize a particular coherent phonon mode of the SWNT with a wide
frequency range in the THz regime for nanomechanics applications, such
as high-Q resonators~\cite{eichler11-nature}, as well as quantum
information science~\cite{ruskov12-compcp,li12-spincp}.  However, we
have to select a single coherent phonon mode with a well-defined
frequency and suppress other phonon modes that have been excited,
which is not possible to achieve by using only a single pulse of
laser.

One possible technique to selectively excite a particular phonon mode
is by using multiple pulses with a determined repetition rate in
pump-probe spectroscopy.  Since two decades ago, several experimental
studies have used femtosecond pulse sequences for optical manipulation
of molecular motion~\cite{weiner90-pulse}, for enhancement of optical
phonons in mixed crystals~\cite{hase98-pulse}, and also for
controlling lattice dynamics at metal
surfaces~\cite{watanabe05-pulse}.  In the case of SWNTs, Kim \emph{et
  al.} have recently shown that, by introducing pulse
shaping~\cite{kim09-cpprl}, multiple pulses with the different
repetition rates that are matched to some radial breathing mode (RBM)
frequencies (around $5.5$--$7.5$ THz) can be used to excite the RBMs
of specific SWNTs in a bundled SWNT sample.  The bundled SWNT sample
consists of several tube species having different chiralities,
represented by a set of integers $(n,m)$, which give the SWNT
circumferencial lengths and diameters~\cite{saito98-phys}.  Since the
RBM frequency is inversely proportional to the SWNT
diameter~\cite{dresselhaus05-raman}, different RBM frequencies will be
excited at the same time if we only use a single laser pulse without
pulse shaping.  By applying the pulse train with a frequency matching
to a RBM frequency of a single chirality SWNT, it is possible to
excite the RBM of the specified SWNT in the bundled
sample~\cite{kim09-cpprl}.

To our knowledge, it has not been shown yet whether or not the
pulse-train technique can be used to generally excite a specific
coherent phonon mode other than the RBMs in SWNTs, for example, the
G-bands in the SWNTs, which have higher frequency ($\sim$$47.7$ THz).
Moreover, it is also interesting to investigate whether or not we can
selectively generate a particular coherent phonon mode for a
single-chirality SWNT which is mixed in the bundled sample.  Due to
the classical nature of coherent phonon
oscillations~\cite{zeiger92-cp}, either in an enriched
single-chirality sample or mixed bundled sample, one might think that
it is trivial to selectively excite a specific phonon mode just by
using a pulse train repetition frequency which is matched to the
phonon frequency.  However, we will show in this paper that, based on
both numerical and analytical methods, the frequency matching is not
the only condition to obtain the selective coherent phonon generation.
Further considerations such as the dependence of phonon selectivity on
the number of pulses and on frequency ratio of different phonon modes
are also discussed.

This paper is organized as follows.  In section~\ref{sec:method}, we
overview the way to calculate the coherent phonon amplitudes and
spectra.  We basically follow the methods described in our previous
papers~\cite{sanders13-review,kim13-cp}, which now takes into account
several laser pulses in the pulse train.  Besides the numerical
method, we also show a possible description for the selective phonon
excitation from the analytical solution of a driven force oscillator,
which is given in section~\ref{sec:analytic}.  In
sections~\ref{sec:results}, particularly in~\ref{sec:rbm} and
\ref{sec:g}, we discuss how we can selectively excite the radial
breathing mode (RBM) and the G-band for a given SWNT by considering
some adjustable external parameters such as the laser pulse width and
the pulse repetition rate (or repetition period).  We examine some
conditions in which each of the phonon modes can be either kept
surviving or suppressed.  In~\ref{sec:relax}, we briefly discuss the
results the effects of carrier relaxation on coherent phonon
generation.  Conclusion and perspective are given in
section~\ref{sec:conclude}.

\section{Methods}
\subsection{Numerical simulation}
\label{sec:method}
We define a coherent phonon mode with wavevector $q = 0$ ($\Gamma$
point phonon) whose amplitude $Q_m$ satisfies a driven oscillator
equation~\cite{stanton94-cpmethod,hu96-qcp,merlin97-cp}
\begin{equation}
\label{eq:drive}
\frac{\partial^2 Q_{m}(t)}{\partial t^2} +
\omega^2_m Q_{m}(t) = S_m(t) ,
\end{equation}
where $m$ labels the phonon mode and $\omega_m$ is the phonon
frequency.  Equation~(\ref{eq:drive}) is solved subject to the initial
conditions $Q_m(0) = 0$ and $\dot{Q}_m(0) = 0$.  The driving force
$S_m(t)$ in the right hand side of equation~(\ref{eq:drive}) is given
by
\begin{equation}
S_m(t) = -\frac{2\omega_m}{\hbar} \sum_{nk} {\cal M}^m_{n}(k) \left[
  f_{n}(k,t) - f^0_{n}(k) \right],
\label{eq:force}
\end{equation}
where $f_{n}(k,t)$ is the time-dependent electron distribution
function and $f^0_{n}(k)$ is the initial equilibrium electron
distribution function before pump pulse is applied.  The derivation
and justification of equations~(\ref{eq:drive}) and (\ref{eq:force})
is provided, for example, in reference~\cite{stanton94-cpmethod}.
Here $n$ labels an electronic state, while $k$ gives the electron
wavevector.  The electronic states of a SWNT are calculated within the
extended tight-binding (ETB) approximation~\cite{popov04-opt}.  The
electron-phonon matrix element ${\cal M}^m_{n}(k)$ is a simplified
notation for ${\cal M}^{m,0}_{nk;nk}$, where
${\cal M}^{m,q}_{n'k';nk}$ is the electron-phonon matrix element in
the ETB model with phonon wavevector $q = k - k'$ and with a
transition from an electronic band $n$ to $n'$~\cite{jiang05-elph}.
It should be noted that in equation~\ref{eq:drive}, we neglect the
phonon damping term which originates from the
anharmonicity~\cite{kim13-cp}.  We expect that if the phonon damping
is included, an additional damping term modifying the coherent phonon
amplitude will appear due to the dephasing time of the coherent
phonon, which could then modulate the coherent phonon amplitude.
However, looking at some experimental measurements on the phonon
dephasing in SWNTs by single pulse pump-probe spectroscopy we estimate
that even the G-band, which has a high frequency, can survive up to
$5\unitps$~\cite{kim12-deph}.  Therefore, we expect that the phonon
damping is not a key factor contributing to the selective phonon
excitation.

In ultrafast pump-probe spectroscopy, a pump laser pulse should have
much shorter pulse width ($\sim$$10\unitfs$) compared with a typical
coherent phonon period ($\sim 20$--$100\unitfs$).  The observed
coherent phonon intensity is proportional to the power spectrum of the
oscillations of optical properties~\cite{zeiger92-cp}.  We see in
equation~\ref{eq:force} that the driving force $S_m(t)$ depends on the
photoexcited electron distribution functions, which can be calculated
by considering the photogeneration of electrons.  The photogeneration
rate from Fermi's golden rule is obtained as~\cite{chuang95}:
\begin{eqnarray}\label{eq:rate}
  \nonumber 
  \frac{\partial f_{n}(k,t)}{\partial t} =&&
  \frac{8 \pi^2 e^2 \ u(t)}{\hbar \ n_g^2 \ (E_{\rm pump})^2}
  \left(\frac{\hbar^2}{m_0} \right) \sum_{n'}
  \left| P_{n n'}(k,t) \right|^2
  \\ &&
  \nonumber \times \Big( f_{n'}(k,t) - f_{n}(k,t) \Big)
  \\ &&
  \times
  \delta \Big( E_{n n'}(k,t) - E_{\rm pump} \Big) ,
\end{eqnarray}
where $E_{n n'}(k,t) = \arrowvert E_{n}(k,t) - E_{n'}(k,t) \arrowvert$
are the $k$ dependent transition energies for a given SWNT at time $t$
of phonon oscillation, $E_{\rm pump}$ is the pump laser energy, $P_{n
  n'}(k,t)$ is the optical matrix element obtained within the dipole
approximation~\cite{alex03-opt,popov04-opt}, and $u(t)$ is the
time-dependent energy density of the pump pulse. It should be noted
that $E_{n n'}(k,t)$ and $P_{n n'}(k,t)$ are modulated by $Q_m(t)$ of
equation~(\ref{eq:drive}) through the change of the C--C bond
length~\cite{kim09-cpprl,nugraha15-cp}.

The energy density of pump pulse, $u(t)$, is related with the pump
fluence $F$ by a relation $F = (c/n_g) \int u(t) dt$, in which $u(t)$
is also assumed to be a Gaussian.  We define a single Gaussian pulse
with an index $j$ as
\begin{equation}
\label{eq:pulse}
u_j(t) = A_p e^{-4 (t-t_j)^2 \ln 2 /2\tau_p^2} ,
\end{equation}
where $A_p = (2 n_g F \sqrt{\ln 2 / \pi})/ (c\tau_p)$, with $c$ is the
speed of light, $\tau_p$ is the laser pulse width, and $t_j$ is the
time at which the peak of the Gaussian is located.  We set the fluence
$F=10^{-5}~{\rm J\ cm}^{-2}$ and refractive index $n_g = 1$.  For the
multiple Gaussian pulses constituting the pulse train, we can sum up
several $u_j(t)$ as follows:
\begin{eqnarray}
\label{eq:sumpulse}
\nonumber u(t) &&= \frac{1}{N_p} \sum_{j=0}^{N_{\rm pulse} - 1} u_j(t)
\\ &&= \frac{1}{N_p} \sum_{j=0}^{N_{\rm pulse} - 1} A_p e^{-4
  (t-t_j)^2 \ln 2 /2\tau_p^2},
\end{eqnarray}
where $N_{\rm pulse}$ is the number of pulses and $t_j$ can be
expressed as $t_j = j T_{\rm rep}$ with $T_{\rm rep}$ is the pulse
repetition period, defined by the time interval between two
neighboring Gaussian pulses.  The normalization constant, $N_p$,
appearing in equation~\ref{eq:sumpulse} is taken into account when we
assume that the pulse train is created by dividing a single laser
pulse into $N_{\rm pulse}$.  In the simplest case, $N_p = N_{\rm
  pulse}$, i.e., the source laser pulse is equally divided into
$N_{\rm pulse}$ Gaussians.  However, if we consider another pulse
shaping, we generally have to integrate the pulse train laser density
so that it has the same laser fluence as the source laser pulse.

We can also consider carrier relaxation or decay process in addition
to the photogeneration process.  Some experimental works have reported
that the photoexcited electrons and holes in SWNTs are nonradiatively
recombined via multiphonon emission processes within $1 \unitps$ time
scale, which is one order of magnitude slower than in
graphite~\cite{ichida02-dynamics,lauret03-carrier}.  We can take these
effects into account by adding a phenomenological term to the
photogeneration rate in equation~(\ref{eq:rate}) as follows:
\begin{equation}
\label{eq:relax}
\left[\frac{\partial f_n(k,t)}{\partial t}\right]_r = -\frac{1}{\tau_r}
[f_n(k,t) - f_n^0(k)], 
\end{equation}
where $\tau_r$ is the phenomenological relaxation time.  When it is
necessary to consider the carrier relaxation effects, we can add the
relaxation term of the form of equation~(\ref{eq:relax}) to the photogeneration
rate in equation~(\ref{eq:rate}) and solve them as a whole. 

In coherent phonon spectroscopy, a probe pulse is used to measure the
absorption coefficient of the SWNT, $\alpha (t)$, at a probe energy
$E_{\rm probe}$ and time $t$:
\begin{eqnarray}
\label{eq:alfa}
\alpha(E_{\rm probe},t) \propto&& \sum_{n n'} \int dk |P_{n n'}(k,t)|^2
\left[f_{n}(k,t) - f_{n'}(k,t) \right] \nonumber\\
&&\times \delta \left[E_{n n'}(k,t) - E_{\rm probe}\right].
\end{eqnarray}
In this process, excitation of coherent phonons by the laser pump
affects the optical properties of the SWNTs, which gives rise to the
modulation of absorption coefficient $\Delta \alpha$.  It should be
noted that, in the numerical calculation, the delta functions in
equations~(\ref{eq:rate}) and (\ref{eq:alfa}) are approximated by a
Lorentzian line shape.  The expression of $\Delta \alpha$ is given by
\begin{equation}
  \Delta \alpha (E_{\rm probe}, t) = -[\alpha (E_{\rm probe}, t) -
  \alpha (E_{\rm probe}, -\infty)],
\end{equation}
where $t=-\infty$ corresponds to the absence of pump pulse.  The
coherent phonon signal is represented by intensity $I$, which
corresponds to the Fourier power spectrum of $\Delta \alpha$ at a
given energy $E_{\rm probe}$:
\begin{equation}
\label{eq:intensity}
I(\omega) = \int e^{-i\omega t} \left|\Delta \alpha (E_{\rm probe},t)
\right|^2 dt,
\end{equation}
where $\omega$ denotes the phonon frequency contributing to the
coherent phonon spectra.

\subsection{Analytical solution}
\label{sec:analytic}

By some approximations, it is possible to derive an analytical
solution to the driven force oscillator considering the pulse train.
Firstly, we start with equation~(\ref{eq:drive}) using the following
simplified notation,
\begin{equation}
\label{eq:harmonik}
\frac{\partial^2 Q(t)}{\partial t^2} + \omega_0^2 Q(t) = S(t),
\end{equation}
where $\omega_0$ is the phonon frequency, and here we have omitted the
index $m$ from equation~(\ref{eq:drive}).  We can solve
equation~(\ref{eq:harmonik}) by using Green's function method which
results in the following integration problem:
\begin{equation}
\label{eq:qt}
Q(t)=\int{d\omega \int{dt'
    \frac{1}{\omega_0^2-\omega^2}e^{i\omega(t-t')}S(t') dt'}}.
\end{equation}

Determining the shape of $S(t)$ is now important to explain the
phenomena of coherent phonon oscillations.  It has been known from
earlier studies that coherent phonon oscillations may be driven by a
displacive force or impulsive force, where the basic shape of the
driving force is close to a step function or delta function,
respectively~\cite{zeiger92-cp,merlin97-cp}.  In the case of SWNTs,
the driving force from photoexcited carriers interacting with SWNT
lattices has been proposed to be displacive, i.e., the shape is close to a
step function with a broadening factor determined by the laser pulse
width $\tau_p$~\cite{nugraha15-cp}.  When we consider the pulse train,
we thus expect that the driving force will be like a staircase, where
there are some steps originating from the number of pulses $N_{\rm
  pulse}$.  It is noted that solving the driven oscillator problem by
combining $\tau_p$ and $N_{\rm pulse}$ for the staircase function is
numerically feasible.  However, an exact analytical solution is quite
complicated to obtain.  In this case, we assume that $\tau_p$ can be
omitted from the driving force for simplicity.  On the other hand, for
the impulsive driving force, we may approximate the shape of the
driving force by Gaussian function.  where it is possible to obtain an
exact analytical solution for $Q(t)$ even by combining $\tau_p$ and
$N_{\rm pulse}$ in the driving force.  In this section, we derive
analytical $Q(t)$ solutions for both cases.  We will show later in the
discussion section that the two solutions can be equivalent when we
focus on the maximum amplitudes of the oscillations, which implies
that the presence of the pulse train with $N_{\rm pulse}$ does not
care on the type of the force, whether it is displacive or impulsive.

For the displacive force, we assume the following form of the
individual driving force component,
\begin{equation}
\label{eq:stdisplacive}
S_D(t') = A_D \Theta(t'-t_j) e^{-\gamma_r t'},
\end{equation}
where $A_D$ is the displacive driving force amplitude,
$\gamma_r=1/\tau_r$ is the relaxation rate parameter, and
$\Theta(t'-t_j)$ is the Heaviside step function with nonzero values
(equal to unity) starting from $t_j$.  By inserting
equation~(\ref{eq:stdisplacive}) into~(\ref{eq:qt}) and using the
residue theorem, the $Q(t)$ solution for one pulse of the displacive
force is obtained as
\begin{eqnarray}
\label{eq:qtdisplacive}
\nonumber Q_D(t) = &&2\pi A_D \Bigg[\frac{e^{-\gamma_r t_j}}
  {\omega_0(\gamma_r^2-\omega_0^2)} \Big(\gamma_r
  \sin{\omega_0(t-t_j)} \\ &&- \omega_0 \cos
      {\omega_0(t-t_j)}\Big)+\frac{e^{-\gamma_r
          t}}{\gamma_r^2+\omega_0^2}\Bigg].
\end{eqnarray}
The full $Q_D(t)$ solution considering $N_{\rm pulse}$ with
substituting $\omega_0 = 2\pi/T_0$ and $t_j = j T_{\rm rep}$ is given
by
\begin{eqnarray}
\label{eq:qtndisplacive}
\nonumber &&Q_D(t) = \\ \nonumber &&2\pi A_D \sum_{j=0}^{N_{\rm
    pulse}-1} \Bigg[\frac{T_0 e^{-\gamma_r j T_{\rm
        rep}}}{2\pi(\gamma_r^2-(\frac{2\pi}{T_0})^2)} \Big\{\gamma_r
  \sin{\big(\frac{2\pi}{T_0}(t-j T_{\rm
      rep})\big)}\\ &&-\frac{2\pi}{T_0} \cos{\big(\frac{2\pi}{T_0}
    (t-j T_{\rm rep})\big)}\Big\} +\frac{e^{-\gamma_r
      t}}{\gamma_r^2+(\frac{2\pi}{T_0})^2}\Bigg],
\end{eqnarray}
where $T_0$ is the phonon period of a particular mode.

For the impulsive force, we assume that $S(t')$ in
equation~(\ref{eq:qt}) can be represented by the Gaussian function
with a peak at $t_j$ and pulse width $\tau_p$.  The individual
impulsive driving force can be written as
\begin{equation}
\label{eq:stimpulsive}
  S_I(t') = A_I e^{-(t'-t_j)^2/\tau_p^2}e^{-\gamma_r t'},
\end{equation}
where $A_I$ is the impulsive driving force amplitude.  Inserting
equation~(\ref{eq:stimpulsive}) into~(\ref{eq:qt}) and solving the
differential equation, $Q(t)$ for one pulse of this impulsive force is
obtained as
\begin{eqnarray}
\label{eq:qtend}
Q_I(t)&=&\frac{2A_I\pi^{3/2}\tau_p}{\omega_0}e^{-[\gamma_r
    t_j+(\omega_0^2-\gamma_r^2)\tau_p^2/4]}\nonumber\\
&&\times\sin[{\omega_0(t-t_j+\gamma_r\tau_p^2/2)}] .
\end{eqnarray}
For $N_{\rm pulse}$ Gaussians, and also substituting $\omega_0 = 2\pi
/ T_0$ and $t_j = j T_{\rm rep}$, we obtain
\begin{eqnarray}
\label{eq:qtN}
Q_I(t) &=& A_I\sqrt{\pi}\tau_p T_0\sum_{j=0}^{N_{\rm pulse} - 1}
e^{-[\gamma_r jT_{\rm
      rep}+((2\pi/T_0)^2-\gamma_r^2)\tau_p^2/4]}\nonumber\\ &&\times\sin
\left[{\frac{2\pi}{T_0} (t - jT_{\rm rep}+\gamma_r\tau_p^2/2)}\right],
\end{eqnarray}
We will use equations~(\ref{eq:qtndisplacive}) and~(\ref{eq:qtN}) to
analyze numerical results in the next section.

\section{Results and discussion}
\label{sec:results}

In the following calculations, we focus to discuss the results for a
$(11,0)$ SWNT.  This chirality is chosen in this work due to its less
expensive calculation cost compared with other chiralities having a
larger number of atoms in the unit cell.  Yet, the $(11,0)$ SWNT is
quite representative for understanding more general results.  We set a
certain laser energy $E_{\rm pump} = E_{\rm probe} = 1.75\unitev$,
which is near the second optical transition energy $E_{22}$ of the
$(11,0)$ SWNT.  The light polarization is fixed parallel to the SWNT
axis.  The RBM phonon period ($T_{\rm RBM}$) for the $(11,0)$ SWNT is
about $111.5\unitfs$, while the G-band period ($T_{\rm G}$) is about
$20.9\unitfs$.  We firstly neglect the the carrier relaxation effects
($\gamma_r = 0$) in the discussions in sections~\ref{sec:rbm}
and~\ref{sec:g} for simplicity.  Then, we discuss the results
including the relaxation effects in section~\ref{sec:relax} to
understand what properties will be affected by the carrier relaxation.

\subsection{RBM selective excitation}
\label{sec:rbm}
\begin{figure}[t!]
  \centering\includegraphics[clip,width=8cm]{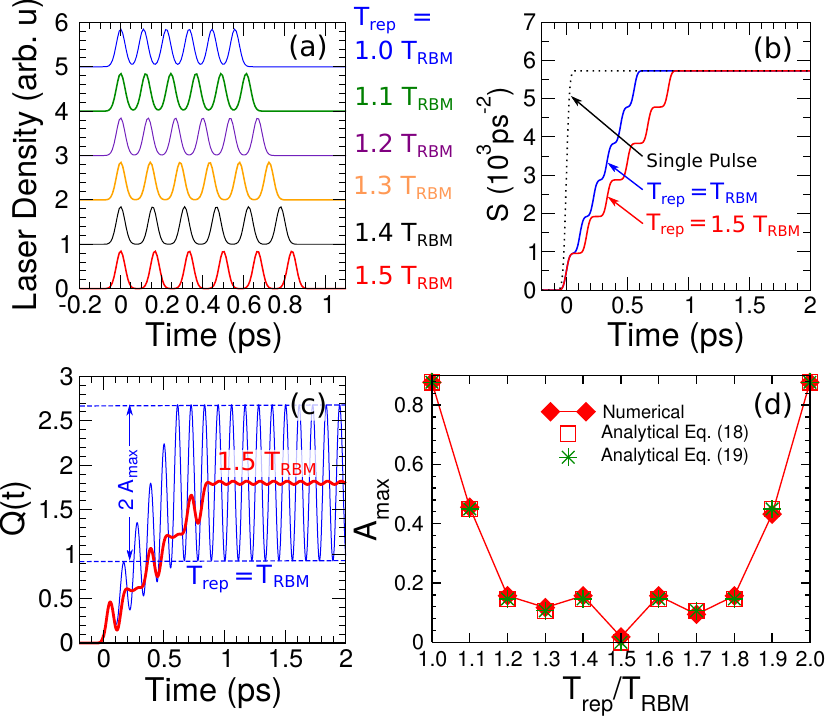}
  \caption{\label{fig1} (a) Laser energy density $u(t)$ for several
    different pulse train repetition periods $T_{\rm rep}$
    ($1.0$--$1.5~T_{\rm RBM}$) which are used to generate the RBM in
    an $(11,0)$ SWNT.  Here $N_{\rm pulse} = 6$ and $T_{\rm RBM}$ for
    the $(11,0)$ SWNT is about $111.5\unitfs$.  For each pulse train,
    we consider $\tau_p = 50\unitfs$. (b) The driving force $S(t)$ due
    to laser pulse trains in (a), shown only for the case of $T_{\rm
      rep}= T_{\rm RBM}$ and $T_{\rm rep} = 1.5~T_{\rm RBM}$.  The
    driving force in the case of a single pulse (not pulse train) is
    given in dotted line for comparison. (c) The resulting coherent
    phonon amplitudes $Q(t)$ generated by $S(t)$ in (b), shown for the
    case of $T_{\rm rep}= T_{\rm RBM}$ and $T_{\rm rep} = 1.5~T_{\rm
      RBM}$.  We additionally define a maximum amplitude $A_{\rm
      max}$, which is shown in panel (c) for $T_{\rm rep} = T_{\rm
      RBM}$ as an example.  (d) The maximum amplitude for each $Q(t)$
    as a function of pulse train repetition period. Filled diamonds
    are from the numerical calculation, while open squares and stars
    are from two different analytical approaches.  The analytical
    $A_{\rm max}$ values are normalized to the numerical value of
    $A_{\rm max}$ at $T_{\rm rep} = T_{\rm RBM}$.}
\end{figure}

In figure~\ref{fig1}, we show the calculated results for the RBM mode
selection in the $(11,0)$ SWNT.  Figure~\ref{fig1}(a) shows the pulse
train consisting of six Gaussian pulses $(N_{\rm pulse} = 6)$ for
several different repetition periods $T_{\rm rep}$.  The repetition
period $T_{\rm rep}$ in figure~\ref{fig1} is expressed as a multiple
of the RBM phonon period, $T_{\rm RBM}$.  For each pulse train, we use
$\tau_p = 50\unitfs$ so that the coherent G-band ($T_{\rm
  G}=20.9\unitfs$) would not be excited.  Example driving forces
$S(t)$ due to laser pulse trains in figure~\ref{fig1}(a) are given in
figure~\ref{fig1}(b), in which we show the case of $T_{\rm rep}=
T_{\rm RBM}$ and $T_{\rm rep} = 1.5~T_{\rm RBM}$, and we also display
the driving force in the case of a single pulse (i.e., not a pulse
train) for comparison.  The driving forces are of displacive type
based on the concept of electron-phonon and electron-photon
interactions developed in section~\ref{sec:method}.
Figure~\ref{fig1}(c) shows the resulting coherent RBM phonon
amplitudes excited by the driving forces from figure~\ref{fig1}(b) for
$T_{\rm rep} = T_{\rm RBM}$ and $T_{\rm rep} = 1.5~T_{\rm RBM}$.  To
check further behavior of the coherent phonon oscillation by changing
the pulse repetition period, we define $A_{\rm max}$ in
figure~\ref{fig1}(c) as the maximum amplitude of the oscillation.  In
figure~\ref{fig1}(d), we show $A_{\rm max}$ for several $T_{\rm rep}$
values within $1.0$--$2.0~T_{\rm RBM}$.

From figure~\ref{fig1}(d), we can see that the numerical value of
$A_{\rm max}$ in the case of $N_{\rm pulse} = 6$ is the largest
(smallest) for $T_{\rm rep} = T_{\rm RBM}$ ($T_{\rm rep} = 1.5~T_{\rm
  RBM}$).  This numerical result can be reproduced analytically using
both equations~(\ref{eq:qtndisplacive}) and (\ref{eq:qtN}) considering
$\gamma_r = 0$ and $T_0 = T_{\rm RBM}$.  We firstly simplify
equation~(\ref{eq:qtndisplacive}) so that it reduces to
\begin{equation}
\label{eq:qtRBM0}
Q_D (t) \propto \sum_{j=0}^{N_{\rm pulse} - 1} \left\{1 +
\cos\left[2\pi\left(\frac{t - j T_{\rm rep}}{T_0}\right)
  \right]\right\} ,
\end{equation}
while equation~(\ref{eq:qtN}) reduces to
\begin{equation}
\label{eq:qtRBM}
Q_I (t) \propto \sum_{j=0}^{N_{\rm pulse} - 1}
e^{-\pi^2\left(\frac{\tau_p}{T_0}\right)^2}
\sin\left[2\pi\left(\frac{t - j T_{\rm rep}}{T_0}\right)\right],
\end{equation}
To compare the analytical solution with the numerical result on the
same plot, we calculate the analytical $A_{\rm max}$ values for
different $T_{\rm rep}$ from equations~(\ref{eq:qtRBM0}) and
(\ref{eq:qtRBM}) normalized to numerical $A_{\rm max}$ at $T_{\rm rep}
= T_0 = T_{\rm RBM}$.  As shown in Fig.~\ref{fig1}(d), the agreement
between the analytical and numerical results is quite excellent.  In
fact, if we perform the summations in equations~(\ref{eq:qtRBM0}) and
(\ref{eq:qtRBM}) explicitly to calculate $A_{\rm max}$ as defined in
figure~\ref{fig1}(c), we find that both equations will have the same
form of $A_{\rm max}$ as follows:
\begin{equation}
\label{eq:amaxanalytic}
A_{\rm max} \propto \left|\frac{\sin[N_{\rm pulse} \pi (T_{\rm rep} /
    T_0)]}{\sin[\pi (T_{\rm rep} / T_0)]} \right|.
\end{equation}
Equation~(\ref{eq:amaxanalytic}) suggests that by using the
pulse-train technique, the behavior of $A_{\rm max}$ is independent of
the detailed shape of the constituting force.  From both numerical and
analytical results, we can also see that the value of $A_{\rm max}$ at
$T_{\rm rep} = T_{\rm RBM}$ is recovered at
$T_{\rm rep} = 2T_{\rm RBM}$, which is also expected for higher
integer-multiple of $T_{\rm RBM}$.  It indicates that as long as the
pulse repetition period matches integer multiple (half-integer
multiple) of the RBM period, we will mostly have the coherent RBM
phonon amplitude enhanced (suppressed).  It should be noted from
equation~(\ref{eq:amaxanalytic}) that the shape of $A_{\rm max}$ as a
function of $T_{\rm rep}/T_{\rm RBM}$ is also
$N_{\rm pulse}$-dependent.  In particular, if $N_{\rm pulse}$ is an
even (odd) number, $A_{\rm max}$ is zero (nonzero but small) at
$T_{\rm rep}$ equals half-integer multiple of $T_{\rm RBM}$.

\begin{figure}[t!]
  \centering\includegraphics[clip,width=8cm]{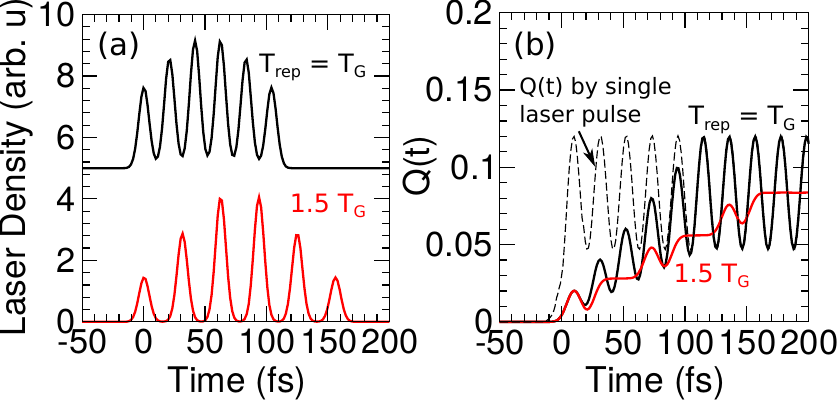}
  \caption{\label{fig2} Selection of the G-band of the (11,0) SWNT by
    a pulse train with $T_{\rm rep}$ varied in terms of $T_{\rm G}$.
    We excite the coherent phonons by laser pulses with $\tau_p =
    10\unitfs$ and $N_{\rm pulse} = 6$.  Panel (a) shows the laser
    density of the pulse train with $T_{\rm rep} = T_{\rm G}$ and
    $T_{\rm rep} = 1.5~T_{\rm G}$, while panel (b) shows the resulting
    coherent G-band phonon amplitude.  For additional comparison, we
    show a dashed line that represents the G-band amplitude generated
    by a single laser pulse.}
\end{figure}

\subsection{G-band selective excitation}
\label{sec:g}
In figure~\ref{fig2}, we present a similar calculation for the G-band
phonon mode, which has a higher-frequency (shorter period) than the
RBM.  In this case, $T_{\rm G}$ for the $(11,0)$ SWNT is about
$20.9\unitfs$.  Morover, to consider an experimental condition in
which a Gaussian pulse shaping (or an envelope function for multiple
pulses) is used to obtain a narrow spectral shape~\cite{kim09-cpprl},
we multiply $u(t)$ with a Gaussian function having a width of half of
the total pulse duration.  We then normalize the new $u(t)$ so as the
pump fluence $F$ is the same as in the case of using a single Gaussian
pulse.  Figure~\ref{fig2}(a) shows the pulse train profile consisting
of six Gaussian pulses with the pulse width $\tau_p = 10\unitfs$.  The
pulse train in figure~\ref{fig2}(a) is varied with two different
repetition periods: $T_{\rm rep} = T_{\rm G}$ and $T_{\rm rep} =
1.5~T_{\rm G}$.  In figure~\ref{fig2}(b), we show the resulting
coherent phonon amplitude by the two cases of pulse train, which is
also compared with the amplitude excited by a single Gaussian pulse.
It can be seen that the pulse repetition period which matches to the
integer multiple (half-integer multiple) of the G-band period will
mostly enhance (suppress) the G-band phonon amplitude.  However, we
find that actually for the higher-frequency coherent phonon modes such
like the G-band (which has smaller amplitudes than the RBM), the
repetition rate with the value of the integer multiple of the G-band
period is \emph{not always} sufficient to completely suppress the RBM
amplitude although the G-band amplitude is already at its possible
largest value.

\begin{figure}[t!]
  \centering\includegraphics[clip,width=75mm]{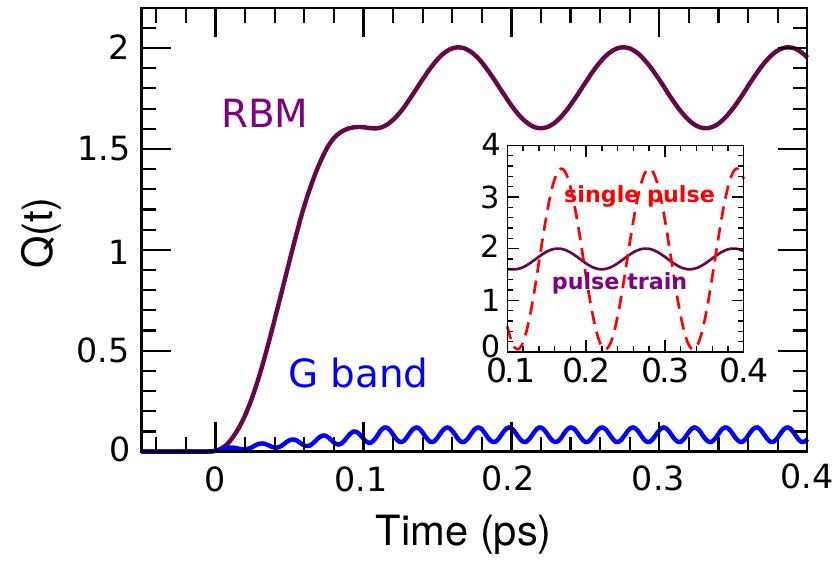}
  \caption{\label{fig3} The coherent RBM and G-band phonon amplitudes
    for the (11,0) SWNT excited by the same pulse train consisting of
    six Gaussian pulses with $\tau_p = 10\unitfs$ and $T_{\rm rep} =
    T_{\rm G} = 20.9\unitfs$.  Inset shows comparison between the RBM
    phonon excited by a single Gaussian pulse (dashed line) and that
    by the pulse train with $N_{\rm pulse} = 6$ (solid line) taken
    from the main figure.}
\end{figure}

In figure~\ref{fig3}, we show a comparison between the coherent RBM
and G-band phonon amplitudes excited with $\tau_p = 10\unitfs$ and
$T_{\rm rep} = T_{\rm G} = 20.9\unitfs$ laser pulses.  We can see that
the RBM still has a larger amplitude compared with the G-band excited
by the same pulse train.  We expect that the origin of this behavior
is due to the dependence of the phonon amplitude on the phonon
frequency, in which $Q_m(t)$ is inversely proportional to $\omega_m$
upon solving the equation~(\ref{eq:drive}), i.e., the higher-frequency
phonon modes tend to have a smaller amplitude than the lower frequency
modes~\cite{nugraha15-cp}, as is also indicated by
equations~(\ref{eq:qtdisplacive}) and (\ref{eq:qtend}).  We note that
the choice of $T_{\rm rep} = T_{\rm G}$ for the case shown in
figure~\ref{fig3} still does not completely give a destructive
interference for the RBM.  To suppress the RBM while still keeping the
G-band, we propose a trick that the repetition period should instead
be of half-integer multiple of the RBM period by keeping the same
number of pulses (e.g., $N_{\rm pulse} = 6$).  Thus, the RBM is
completely suppressed because of destructive interference, while the
G-band still survives.  Another alternative is by still using
$T_{\rm rep} = T_{\rm G}$ but increasing the number of pulses.  This
latter way may be justified by checking the behavior of a certain
phonon mode amplitude excited with different number of pulses.  For
example, in the inset of figure~\ref{fig3} we show RBM phonon
amplitudes generated by a single Gaussian pulse (dashed line) and by a
pulse (solid line) with the same $\tau_p = 10\unitfs$ (additionally we
also have $N_{\rm pulse} = 6$ and
$T_{\rm rep} = T_{\rm G} = 20.9\unitfs$ for the pulse train).  We can
see that using the pulse train with $T_{\rm rep} = T_{\rm G}$ can
already reduce the RBM amplitude because the repetition period does
not coincide with the integer multiple of $T_{\rm RBM}$.

\begin{figure}[t!]
  \centering\includegraphics[clip,width=8cm]{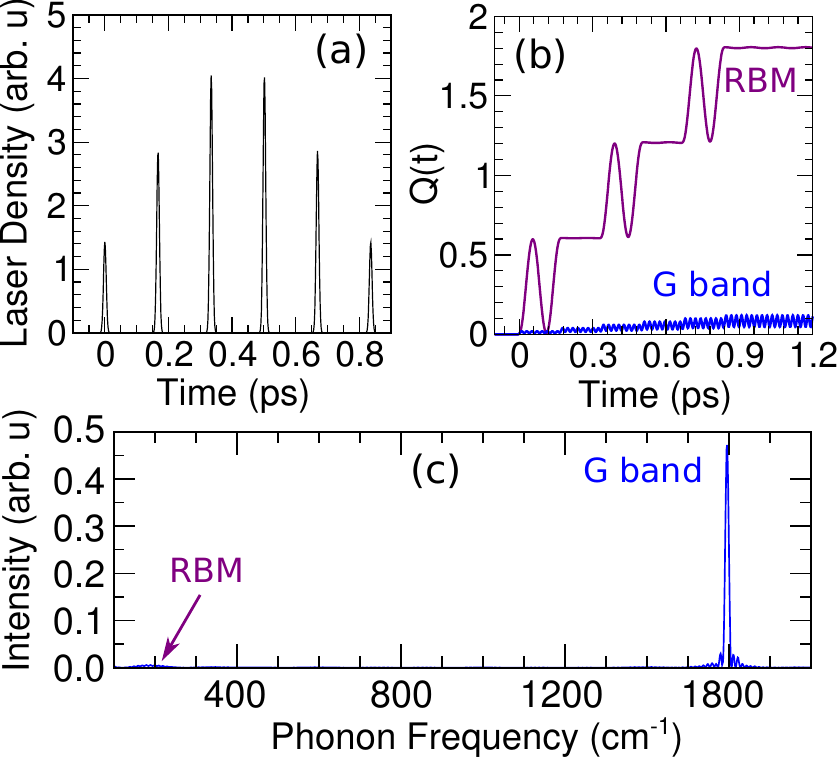}
  \caption{\label{fig4} A possible way to suppress the RBM while at
    the same time keeping the G-band is by using $T_{\rm rep}$ with
    the value of half-integer multiple of $T_{\rm RBM}$, but $\tau_p$
    should be less than $T_{\rm G}$.  Panel (a) shows the laser
    density of the pulse train with $T_{\rm rep}=1.5~T_{\rm RBM}$,
    $\tau_p = 10\unitfs$, and $N_{\rm pulse}=6$.  Panel (b) shows the
    resulting RBM and G-band oscillation amplitudes.  Panel (c) shows
    the coherent phonon intensity for the RBM and G-band.  The
    intensity is the Fourier transform of the absorption modulation
    $\Delta \alpha$ due to the coherent phonon oscillations.  The RBM
    in this case almost disappears while the G-band survives.}
\end{figure}

Now let us see what happens to the RBM and the G-band when
$T_{\rm rep}$ is selected to be half-integer multiple of
$T_{\rm RBM}$.  Figure~\ref{fig4}(a) shows the pulse train profile
with repetition period $T_{\rm rep} = 1.5~T_{\rm RBM}$ and pulse width
$\tau_p = 10\unitfs$ for $(11,0)$ SWNT, while figure~\ref{fig4}(b)
shows the resulting coherent phonon amplitude for the RBM and the
G-band.  We do not consider yet the decay of the photoexcited cariers
(in several $\unitps$) for simplicity.  It is noted in
figure~\ref{fig4}(b) that the displacive constant amplitude of $Q(t)$
shifts for every additional pulse.  It is reasonable in our
approximation because a cancellation of vibration occurs while
expansion of bond length takes place by increasing the number of
photo-excited carriers.  However, the constant value of $Q(t)$ does
not contribute to $I(\omega)$ in equation~(\ref{eq:intensity}) except
for $\omega=0$.  It means that applying $\tau_p = 10\unitfs$ warrants
the excitation of the G-band, while by applying
$T_{\rm rep} = 1.5~T_{\rm RBM}$ we can filter the RBM.  We expect that
the origin of this phenomenon is related to the interference effect.
The RBM is completely suppressed due to the destructive interference.
We further confirm this argument by calculating the coherent phonon
intensity of the RBM and the G-band from
equations~(\ref{eq:alfa})--(\ref{eq:intensity}), as shown in
figure~\ref{fig4}(c). In the experiment of coherent phonon
spectroscopy, the coherent phonon intensity is a Fourier transform
intensity (power spectrum) from either the differential transmittance
or reflectance~\cite{zeiger92-cp,merlin97-cp}.  However, in this work,
we calculate the coherent phonon intensity from the absorption
modulation $\Delta \alpha$ because $\Delta T$ or $\Delta R$ is
proportional to $\Delta \alpha$.  From figure~\ref{fig4}(c), we
clearly see that the RBM (G-band) could be suppressed (kept) by this
technique.  It should be noted that some ripples appearing near the
main peaks are due to the numerical limitation of our Fourier
transform program, which can safely be neglected.

Although we have shown that $T_{\rm rep} = 1.5~T_{\rm RBM}$ could be
used to keep the G-band while suppresing the RBM, this approach might
not be handy for experimentalists since they need to ensure their
sample only consists of a single SWNT chirality.  To overcome this
limitation for the bundled sample having a lot of SWNT chiralities
mixed inside it, we may employ another approach of pulse-train
treatment by utilizing many pulses within the pulse train.  In this
case, if we want to selectively excite the G-band, it is not necessary
to apply $T_{\rm rep} = 1.5~T_{\rm RBM}$.  Instead, we may again use
$T_{\rm rep} = T_{\rm G}$ (or integer multiple of the G-band period)
by increasing the number of laser pulses. 

\begin{figure}[t!]
  \centering\includegraphics[clip,width=85mm]{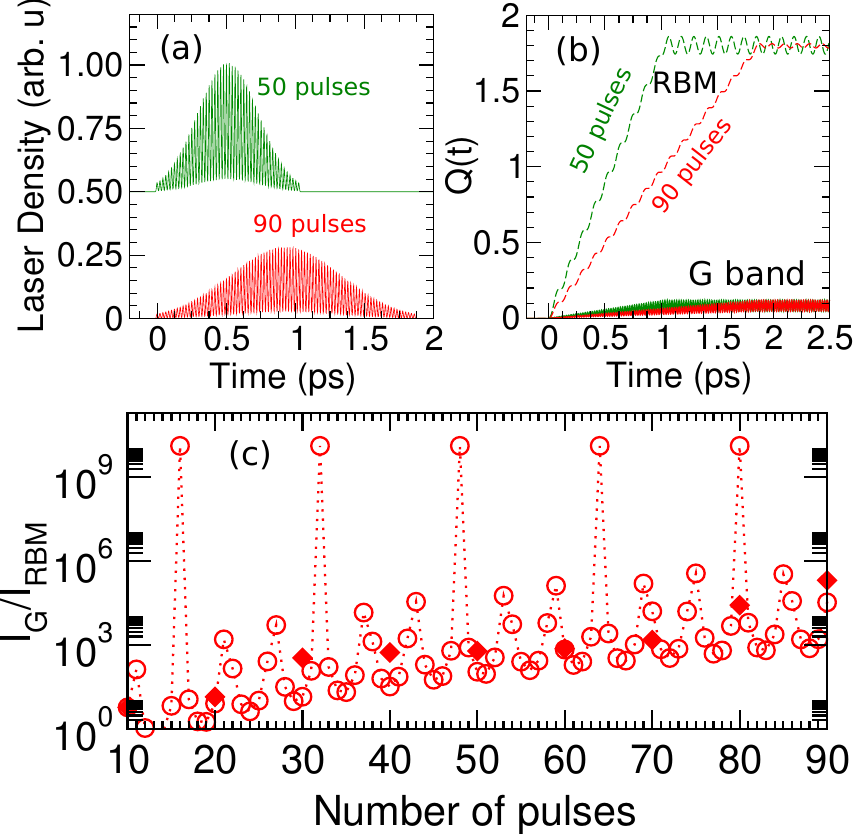}
  \caption{\label{fig5} An alternative approach to keep the G-band and
    suppress the RBM by considering the dependence of coherent phonon
    intensity on the number of pulses in the pulse train.  (a) Laser
    density profile in the case of $50$ and $90$ laser pulses.  (b)
    The resulting coherent phonon amplitudes due to the applied pulse
    train for the RBM (dashed line) and the G-band (solid line), each
    is shown for the two different number of laser pulses.  (c) Ratio
    between coherent phonon intensity of the G band ($I_{\rm G}$) and
    the RBM ($I_{\rm RBM}$) versus the number of pulses.  The
    numerical results are denoted by filled diamonds, plotted every 10
    pulses due to heavy computational time.  The analytical results
    are denoted by open circles, plotted every 1 pulse and connected
    with dotted lines as a guide for eyes.  Note that for comparing
    two approaches, the analytical results are normalized to numerical
    $I_{\rm G}/I_{\rm RBM}$ at $N_{\rm pulse} = 10$.  The SWNT
    chirality considered here is $(11,0)$ and the pulse repetition
    rate is $T_{\rm rep} = T_{\rm G}$.}
\end{figure}

Figure~\ref{fig5}(a) shows an example of the pulse train with $50$
Gaussian pulses and $90$ Gaussian pulses with $\tau_p = 10\unitfs$.
Figure~\ref{fig5}(b) displays the resulting coherent phonon amplitude
for the RBM and G-band of the $(11,0)$ SWNT.  We can see that
basically the larger the number of pulses is, the smaller the RBM
amplitude could be.  To confirm this behavior, we calculate the ratio
between the peak intensity of the G-band and that of the RBM, denoted
by $I_{\rm G} / I_{\rm RBM}$, by varying $N_{\rm pulse}$ within
$10$--$90$.  We perform the calculation by both numerical and
analytical methods and the results are shown in figure~\ref{fig5}(c).
The numerical results are denoted by filled diamonds in
figure~\ref{fig5}(c).  The numerical calculation was performed for
every $10$-pulse increment due to the expensive calculation by
increasing $N_{\rm pulse}$.

We find that as we increase $N_{\rm pulse}$, the value of $I_{\rm G}$
can be much larger than $I_{\rm RBM}$, which indicates a very clear
selectivity of the G-band phonon.  The reason of the fine selection is
that by using larger $N_{\rm pulse}$ with $T_{\rm rep} = T_{\rm G}$,
there is a higher possibility to have destructive interference for the
RBM.  To understand this behavior more clearly, we examine the
analytical solution of $A_{\rm max}$ from
equation~(\ref{eq:amaxanalytic}) for the G-band ($A_{\rm G}$) and for
the RBM ($A_{\rm RBM}$), and then we plot the analytical $I_{\rm
  G}/I_{\rm RBM}$ by assuming that
\begin{equation}
\frac{I_{\rm G}}{I_{\rm RBM}} \propto \left(\frac{A_{\rm G}}{A_{\rm RBM}}\right)^4 .
\end{equation}
This assumption is justified as follows.  We estimate that the
coherent phonon amplitude $Q(t)$ (or $A_{\rm max}$) linearly
influences the optical matrix element $P_{nn^\prime} (k,t)$ in
equation~(\ref{eq:alfa}) which has a power of two, while the coherent
phonon intensity in equation~(\ref{eq:intensity}) also has another
power of two.  Therefore, the intensity can be roughly approximated
proportional to $A_{\rm max}^4$.  By inserting some parameters from
the numerical results ($T_{\rm rep}$, $T_{\rm G}$, and $T_{\rm RBM}$)
into the analytical formula, we can plot the analytical $I_{\rm
  G}/I_{\rm RBM}$ for the increase of $N_{pulse}$ every 1 pulse,
denoted by open circles in figure~\ref{fig5}(c).  To compare the two
approaches, we also normalize the analytical results to the numerical
$I_{\rm G}/I_{\rm RBM}$ at $N_{\rm pulse} = 10$. 

It can be seen that the basic trend of the numerical results in
figure~\ref{fig5}(c) agrees well with the analytical results, i.e.,
increasing $N_{\rm pulse}$ tends to improve the G-band selectivity.
Interestingly, the analytical results in figure~\ref{fig5}(c) give us
more insight on the other regimes of $N_{\rm pulse}$, in which some
sudden increase of $I_{\rm G}/I_{\rm RBM}$ may take place periodically
at particular values of $N_{\rm pulse}$.  We expect that the origin of
this behavior for $T_{\rm rep} = T_{\rm G}$ might be related with the
following relation:
\begin{equation}
\label{eq:magic}
r_i = \frac{N_{\rm pulse} T_{\rm G}}{T_{\rm RBM}} \approx i,
\end{equation}
where $r_{i}$ is a magic ratio and $i$ is a positive integer.  For
example, we can check that for $T_{\rm G} = 20.9\unitfs$ and
$T_{\rm RBM} = 111.5\unitfs$, the values of $r_i$ for
$N_{\rm pulse} = 16, 32$, and $48$ correspond to integers $i = 3, 6$,
and $9$, respectively.  It means that it is still even possible to use
few number of pulses in the pulse train to obtain a good selection of
the G-band if we can precisely determine the ratio of the G-band and
the RBM periods.  Furthermore, we may also increase $T_{\rm rep}$ by
using a larger integer multiple of $T_{\rm G}$ to overcome the
difficulty in making ultrashort repetition periods on the order of
subfemtoseconds.  In that case, the value of magic ratio is scaled by
the integer multiplier.

We expect that the G-band selection method discussed in this section
might be useful if we have many different SWNT chiralities in a
bundled sample, in which we can suppress the RBMs of different
chiralities and keep the G-band.  A remaining difficulty for
experimentalists regarding this technique might be on generating large
number of pulses in the pulse train~\cite{weiner90-pulse}.  For
example, the previous experiment on the RBM selectivity have used
Gaussian pulses up to $30$--$40$
pulses~\cite{kim09-cpprl,sanders09-cp}.  Hopefully more development in
pump-probe spectroscopy in the near future can allow us to realize a
better pulse train system for the G-band and other phonons selective
excitation.

\subsection{Effect of carrier relaxation}
\label{sec:relax}
In the above discussions, we neglected the effect of photoexcited
carrier relaxation since we only focused on the RBM and G-band
selectivity by considering the pulse train repetition period and
number of pulses.  Neglecting the carrier relaxation time is safe
enough for that purpose because the relaxation time $\tau_r~\sim
1~\unitps$ is long enough compared with both the pulse duration and
the phonon period so that it would not alter the conditions of phonon
selectivity.  However, the photoexcited carrier relaxation can affect
the exact amplitude of coherent phonons and hence the coherent phonon
intensity, especially when the number of pulses increase, the total
pulse duration becomes comparable to $\tau_r$.  Here we are
particularly interested in how the coherent phonon intensity would
evolve by changing the number of pulses in the pulse train when we
include the carrier relaxation effects.  The relaxation
effects are considered through equation~(\ref{eq:relax}) and the
coherent phonon intensity is recalculated using
equations~(\ref{eq:alfa})--(\ref{eq:intensity}).

\begin{figure}[t!]
   \centering\includegraphics[clip,width=8cm]{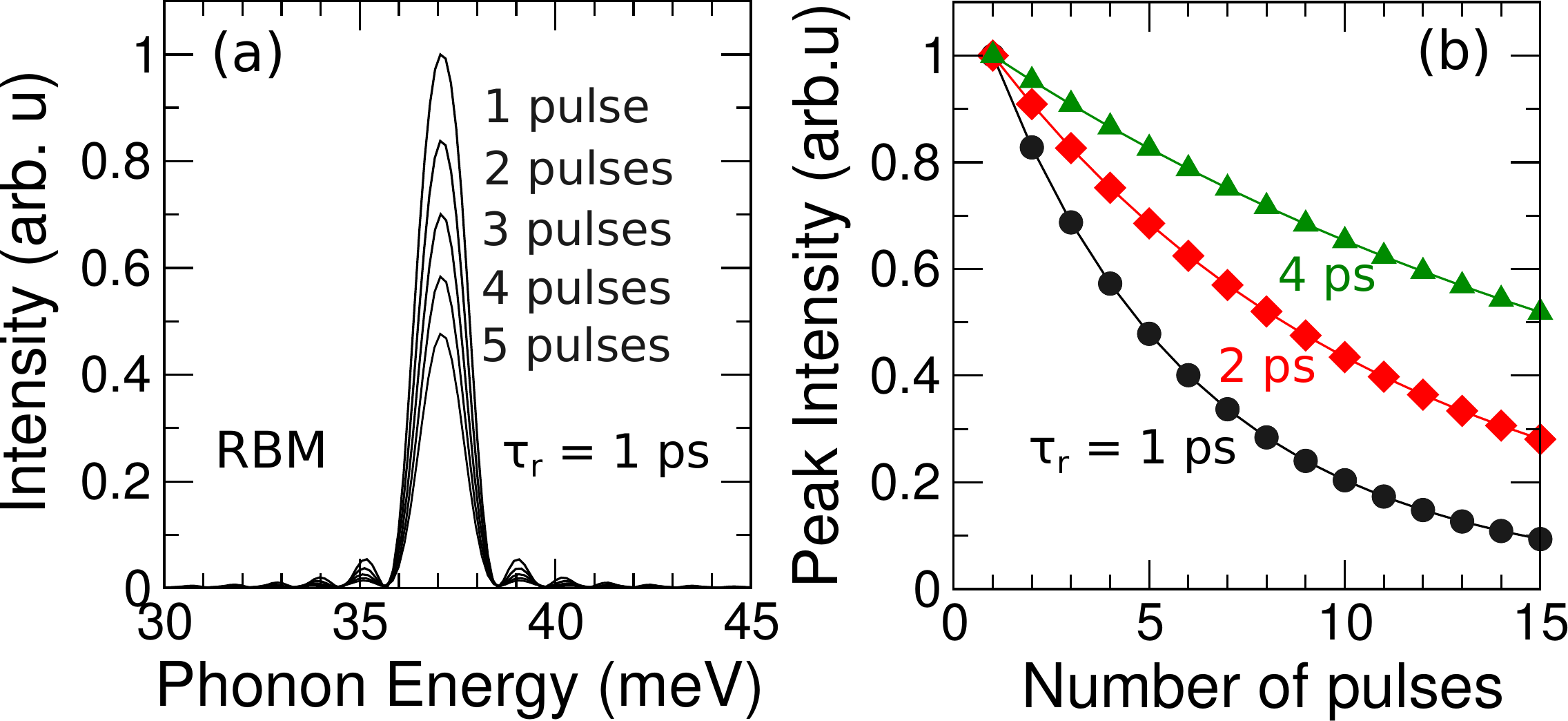}
   \caption{\label{fig6} Carrier relaxation effect.  (a) Coherent RBM
     phonon intensity for the (11,0) SWNT calculated by changing the
     number of pulses.  The relaxation time is set to be $\tau_r =
     1\unitps$.  (b) Comparison of the coherent RBM phonon intensity
     for the (11,0) SWNT for different relaxation time: $\tau_r =
     1\unitps$ (circles), $\tau_r = 2\unitps$ (diamonds), and $\tau_r
     = 4\unitps$ (triangles), which fit well with analytical lines.
     The pulse train repetition period in both panels (a) and (b) is
     $T_{\rm rep} = T_{\rm RBM}$ and the pulse width is $\tau_p =
     50\unitfs$.  All the intensity values are normalized to unity
     with respect to the intensity at $N_{\rm pulse} = 1$.}
\end{figure}

In figure~\ref{fig6}(a), we show an example of the effects of the
carrier relaxation on the coherent phonon intensity, in which the RBM
intensity decreases when we increase the number of pulses in the pulse
train.  In this case, we set the relaxation time $\tau_r = 1\unitps$,
the pulse repetition period $T_{\rm rep} = T_{\rm RBM}$, and the pulse
width $\tau_p = 50\unitfs$.  In figure~\ref{fig6}(b), we vary the
values of $\tau_r$ and plot the peak RBM intensity as a function of
number of pulses for $\tau_r = 1\unitps$, $2\unitps$, and $4\unitps$.
We also fit the numerical data with equation~(\ref{eq:qtN}),
calculating its $A_{\rm max}$ for the RBM with an assumption that
$I_{\rm RBM} \propto (A_{\rm RBM})^4$.  As the relaxation time
increases, a flatter function of peak intensity is obtained as a
function of number of pulses following an exponential feature.  In the
limit of $\tau_r \rightarrow \infty$, the peak intensity becomes
constant, independent of the number of pulses, since we have assumed
that the source laser power is divided into several pulses in the
pulse train.  We expect that such a behavior could be useful to
estimate the carrier relaxation time from experiments if we suppose
that $\tau_r$ is initially unknown, i.e., experimentalists can try
measuring the coherent phonon spectra by changing the number of pulses
in the pulse train and observe the exponential decrease of the peak
intensity versus number of pulses.

\section{Conclusion}
\label{sec:conclude}
We have obtained some conditions for the pulse-train technique to
excite a specific coherent phonon mode in SWNTs while suppressing the
other phonon modes.  As the basic condition, it is necessary to use
the laser pulse width shorter than a typical phonon period for such a
phonon mode to be excited.  Then, for the RBM, the mode selection can
be achieved if the pulse repetition period matches with integer
multiple of the phonon period.  In the case of the G-band, the
repetition period should be of half-integer multiple of the RBM phonon
period or of integer multiple of the G-band period with a large number
of pulses giving a certain value of magic ratio.  From this
simulation, we expect that the pulse-train technique can be generally
used for selectively exciting a specific phonon mode in other
materials, which might be useful for optomechanical applications or
for developing phononic devices.

\section*{Acknowledgments}
We are grateful to Prof.~C.~J.~Stanton, Prof. J. Kono, Dr. J.-H. Kim,
and Dr.~G.~D.~Sanders (University of Florida, USA) for our previous
collaborations which stimulated the present work and for their sharing
the coherent phonon simulation code which we have developed further in
our group. A.R.T.N. acknowledges the Interdepartmental Doctoral Degree
Program for Material Science Leaders at Tohoku University for
providing a financial support.  R.S.  and E.H.H. acknowledge JSPS
KAKENHI Grant Nos. JP25286005 and JP225107005.

\section*{References}


\begin{thebibliography}{10}
\expandafter\ifx\csname url\endcsname\relax
  \def\url#1{{\tt #1}}\fi
\expandafter\ifx\csname urlprefix\endcsname\relax\def\urlprefix{URL }\fi
\providecommand{\eprint}[2][]{\url{#2}}

\bibitem{dumi04}
Dumitric\ifmmode~\u{a}\else \u{a}\fi{} T, Garcia M~E, Jeschke H~O and Yakobson
  B~I 2004 {\em Phys. Rev. Lett.\/} {\bf 92} 117401

\bibitem{gambetta06-cp}
Gambetta A, Manzoni C, Menna E, Meneghetti M, Cerullo G, Lanzani G, Tretiak S,
  Piryatinski A, Saxena A, Martin R~L and Bishop A~R 2006 {\em Nat. Phys.\/}
  {\bf 2} 515--520

\bibitem{lim06-cpexp}
Lim Y~S, Yee K~J, Kim J~H, Haroz E~H, Shaver J, Kono J, Doorn S~K, Hauge R~H
  and Smalley R~E 2006 {\em Nano Lett.\/} {\bf 6} 2696--2700

\bibitem{luer09-cp}
L\"uer L, Gadermaier C, Crochet J, Hertel T, Brida D and Lanzani G 2009 {\em
  Phys. Rev. Lett.\/} {\bf 102} 127401

\bibitem{makino09-cpdoping}
Makino K, Hirano A, Shiraki K, Maeda Y and Hase M 2009 {\em Phys. Rev. B\/}
  {\bf 80} 245428

\bibitem{zeiger92-cp}
Zeiger H~J, Vidal J, Cheng T~K, Ippen E~P, Dresselhaus G and Dresselhaus M~S
  1992 {\em Phys. Rev. B\/} {\bf 45} 768--778

\bibitem{stanton94-cpmethod}
Kuznetsov A~V and Stanton C~J 1994 {\em Phys. Rev. Lett.\/} {\bf 73} 3243--3246

\bibitem{hu96-qcp}
Hu X and Nori F 1996 {\em Phys. Rev. B\/} {\bf 53} 2419--2424

\bibitem{merlin97-cp}
Merlin R 1997 {\em Solid State Commun.\/} {\bf 102} 207--220

\bibitem{lim14-cpfund}
Lim Y~S, Nugraha A~R~T, Cho S~J, Noh M~Y, Yoon E~J, Liu H, Kim J~H, Telg H,
  Hároz E~H, Sanders G~D, Baik S~H, Kataura H, Doorn S~K, Stanton C~J, Saito
  R, Kono J and Joo T 2014 {\em Nano Lett.\/} {\bf 14} 1426--1432

\bibitem{eichler11-nature}
Eichler A, Moser J, Chaste J, Zdrojek M, Wilson-Rae I and Bachtold A 2011 {\em
  Nat. Nanotechnol.\/} {\bf 6} 339--342

\bibitem{ruskov12-compcp}
Ruskov R and Tahan C 2012 {\em JPCS\/} {\bf 398} 012011

\bibitem{li12-spincp}
Li J~J and Zhu K~D 2012 {\em Sci. Rep.\/} {\bf 2} 903

\bibitem{weiner90-pulse}
Weiner A~M, Leaird D~E, Wiederrecht G~P and Nelson K~A 1990 {\em Science\/}
  {\bf 247} 1317

\bibitem{hase98-pulse}
Hase M, Itano T, Mizoguchi K and Nakashima S~i 1998 {\em Jpn. J. Appl. Phys.\/}
  {\bf 37} L281

\bibitem{watanabe05-pulse}
Watanabe K, Takagi N and Matsumoto Y 2005 {\em Phys. Chem. Chem. Phys.\/} {\bf
  7} 2697

\bibitem{kim09-cpprl}
Kim J~H, Han K~J, Kim N~J, Yee K~J, Lim Y~S, Sanders G~D, Stanton C~J,
  Booshehri L~G, H\'aroz E~H and Kono J 2009 {\em Phys. Rev. Lett.\/} {\bf 102}
  037402

\bibitem{saito98-phys}
Saito R, Dresselhaus G and Dresselhaus M~S 1998 {\em Physical Properties of
  Carbon Nanotubes\/} (London: Imperial College Press)

\bibitem{dresselhaus05-raman}
Dresselhaus M~S, Dresselhaus G, Saito R and Jorio A 2005 {\em Phys. Rep.\/}
  {\bf 409} 47--99

\bibitem{sanders13-review}
Sanders G~D, Nugraha A~R~T, Sato K, Kim J~H, Kono J, Saito R and Stanton C~J
  2013 {\em J. Phys. Condens. Mattter\/} {\bf 25} 144201

\bibitem{kim13-cp}
Kim J~H, Nugraha A~R~T, Booshehri L~G, Haroz E~H, Sato K, Sanders G~D, Yee K~J,
  Lim Y~S, Stanton C~J, Saito R and Kono J 2013 {\em Chem. Phys.\/} {\bf 413}
  55--80 ISSN 0301-0104

\bibitem{popov04-opt}
Popov V~N and Henrard L 2004 {\em Phys. Rev. B\/} {\bf 70} 115407

\bibitem{jiang05-elph}
Jiang J, Saito R, Samsonidze G~G, Chou S~G, Jorio A, Dresselhaus G and
  Dresselhaus M~S 2005 {\em Phys. Rev. B\/} {\bf 72} 235408

\bibitem{kim12-deph}
Kim J~H, Yee K~J, Lim Y~S, Booshehri L~G, H\'aroz E~H and Kono J 2012 {\em
  Phys. Rev. B\/} {\bf 86}(16) 161415

\bibitem{chuang95}
Chuang S~L 1995 {\em Physics of optoelectronic devices\/} (New York: Wiley)

\bibitem{alex03-opt}
Gr\"uneis A, Saito R, Samsonidze G~G, Kimura T, Pimenta M~A, Jorio A, Filho
  A~G~S, Dresselhaus G and Dresselhaus M~S 2003 {\em Phys. Rev. B\/} {\bf 67}
  165402

\bibitem{nugraha15-cp}
Nugraha A~R~T, Hasdeo E~H, Sanders G~D, Stanton C~J and Saito R 2015 {\em Phys.
  Rev. B\/} {\bf 91} 045406

\bibitem{ichida02-dynamics}
Ichida M, Hamanaka Y, Kataura H, Achiba Y and Nakamura A 2002 {\em Physica B\/}
  {\bf 323} 237--238

\bibitem{lauret03-carrier}
Lauret J~S, Voisin C, Cassabois G, Delalande C, Roussignol P, Jost O and Capes
  L 2003 {\em Phys. Rev. Lett.\/} {\bf 90} 057404

\bibitem{sanders09-cp}
Sanders G~D, Stanton C~J, Kim J~H, Yee K~J, Lim Y~S, H\'aroz E~H, Booshehri
  L~G, Kono J and Saito R 2009 {\em Phys. Rev. B\/} {\bf 79} 205434

\end{thebibliography}
\providecommand{\newblock}{}

\end{document}